\documentclass{emulateapj} 
\usepackage{apjfonts}
\usepackage{epsfig}

\begin{document}
\slugcomment{ApJ Supplement Series, in press}
\newcommand{\LeeZinn}{\mathcal{L}}

\shortauthors{C\'aceres \& Catelan} 
\shorttitle{The RRL PL Relation in SDSS's $ugriz$}

\title{The Period-Luminosity Relation of RR Lyrae Stars in the SDSS
  Photometric System}

\author{C. C\'{a}ceres \& M. Catelan} 
\affil{Pontificia Universidad Cat\'{o}lica de Chile, Departamento de 
       Astronom\'{\i}a y Astrof\'{\i}sica, \\ Av. Vicu\~{n}a Mackenna 4860, 
      782-0436 Macul, Santiago, Chile; \email{cccacere,mcatelan@astro.puc.cl}
      }

\begin{abstract}
We provide the first detailed study of the RR Lyrae period-luminosity (PL)
relation in the $ugriz$ bandpasses of the Sloan Digital Sky Survey (SDSS) 
filter system. We argue that tight, simple PL relations are not present 
in the SDSS filters, except for the redder bandpasses $i$ and 
(especially) $z$. However, for all bandpasses, we show that, by incorporating
terms involving a (fairly reddening-independent) 
``pseudo-color'' $C_0 \equiv (u-g)_0 - (g-r)_0$, tight 
(non-linear) relations do obtain. We provide theoretically calibrated such
relations in the present paper, which should be useful to derive precise absolute 
magnitudes (hence distances) and intrinsic colors (hence reddening values) 
to even {\em individual} field RR Lyrae stars. For applications to cases where
photometry in all five passbands may not be available, we also provide simple
(though less precise) average PL relations for the $i$ and $z$ bandpasses, 
which read as follows: 

\begin{displaymath}
M_z = 0.839 - 1.295 \, \log P + 0.211\, \log Z,
\end{displaymath}

\begin{displaymath}
M_i = 0.908 - 1.035 \, \log P + 0.220\, \log Z.
\end{displaymath}

\noindent Similarly, simple 
period-color relations for $(r-i)_0$, $(g-r)_0$, and $(u-z)_0$ are 
also provided. 

\end{abstract}

	  
\section{Introduction}
The Sloan Digital Sky Survey (SDSS) has given the scientific comunity 
an unprecedented chance to systematically map a large area of the sky, 
using its own 
special 5-band filter system. In the process, an overwhelming amount 
of data has been ammassed, which can be used to perform many different 
types of scientific studies. Of particular interest to us is the fact 
that the SDSS has also provided, with unprecedented detail, a map of 
the spatial distribution of different stellar populations accross the 
Galaxy, which is increasingly being used to trace new structures in 
the Galactic halo \citep[e.g.,][]{vbea06}, some of which may plausibly 
be related to the hundreds of elusive ``protogalactic fragments'' that 
are predicted, in the $\Lambda$CDM cosmological paradigm, to have given 
birth to a Galaxy like the Milky Way \citep[e.g.,][]{maea03}. 
In this sense, RR Lyrae stars have proven to be a stellar component 
that is consistently present in most, and possibly all, of these 
structures \citep[e.g.,][]{cgea08,ckea08}. Light curves for many variable 
stars in the SDSS filter system have been provided by the SDSS~II 
Survey, a 3-year extension of the original SDSS survey, with images 
of the same fields taken every other night, with the main goal to 
detect supernovae explosions. However, it remains at present 
difficult to obtain reliable distances therefrom, 
since so far no detailed study of the 
properties of RR Lyrae stars in the SDSS system has been performed, 
with the notable exception of \citet{mmea06}. 
In the same vein, it is still not possible at present to directly extract 
reliable distance information from detailed RR Lyrae light curves  
obtained in the SDSS system, as are now becoming increasingly common
\citep*[e.g.,][]{ndlea07,bsea07,rwea08}.


\begin{figure*}[t]
\centering
  \includegraphics*[width=17.5cm]{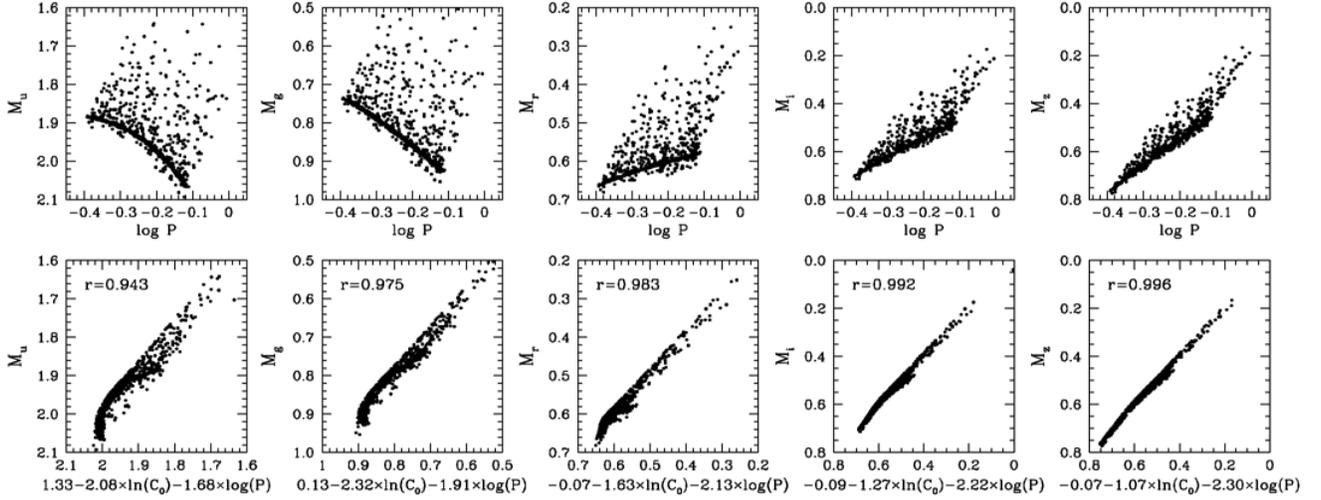}
  \caption{{\em Upper panels}: RR Lyrae PL relations
    for the different indicated SDSS filters. {\em Lower panels}:
    Corresponding RR Lyrae distributions in the absolute
    magnitude--log-period--(pseudo-)color plane. Note the dramatic
    reduction in scatter that is brought about with the inclusion of a
    $C_0$-dependent term (the correlation coefficient $r$ is shown in
    the lower panels). All plots show 650 randomly chosen synthetic RR Lyrae 
    stars from an HB simulation with $Z = 0.002$ and an intermediate HB type.
	}
      \label{fig:GENMAG}
\end{figure*}


\begin{figure*}[t]
\centering
  \includegraphics*[width=17.5cm]{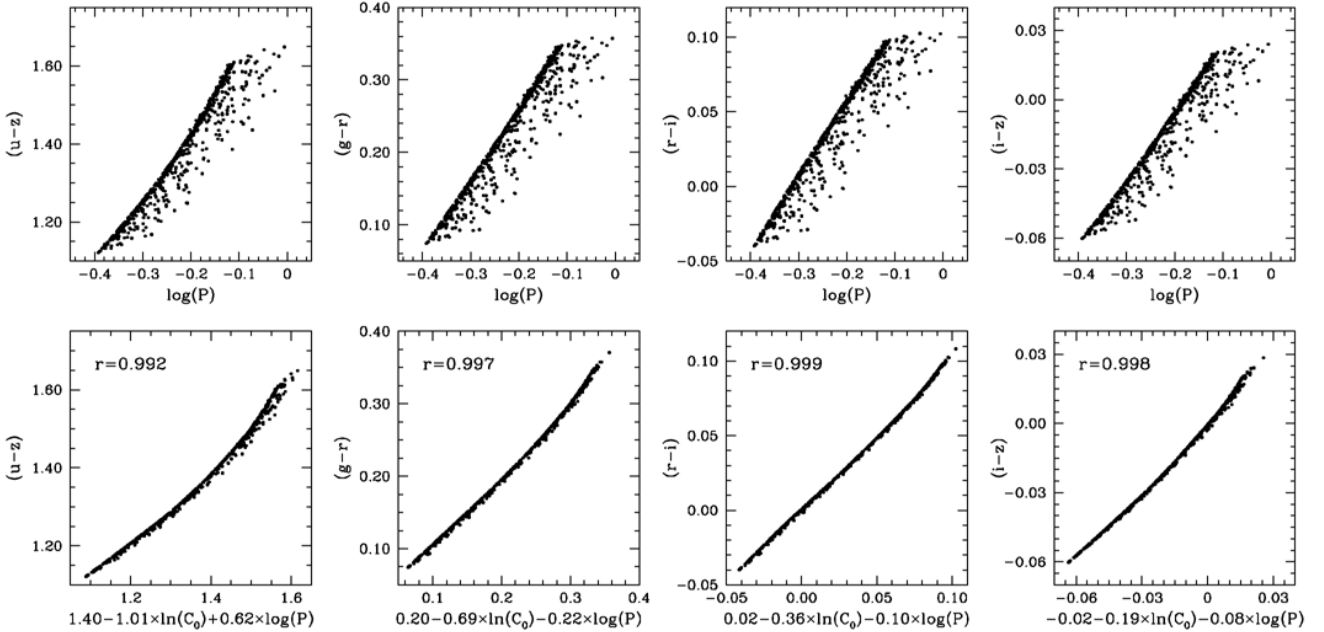}

  \caption{{\em Upper panels}: RR Lyrae PC relations for the
    $(u\!-\!z)$, $(g\!-\!r)$, $(r\!-\!i)$ and $(i\!-\!z)$ SDSS-based
    colors.  {\em Lower panels}: Corresponding RR Lyrae distributions in
    the color--log-period--(pseudo-)color plane. Note the dramatic
    reduction in scatter that is brought about with the inclusion of a
    $C_0$-dependent term (the correlation coefficient $r$ is shown in
    the lower panels).  All plots show 650 randomly chosen synthetic RR Lyrae
    stars from an HB simulation with $Z = 0.002$ and an intermediate HB type.
    }
      \label{fig:GENCOL}
\end{figure*}

Accordingly, the purpose of the present study is to 
provide the first systematic investigation of the RR Lyrae 
period-luminosity (PL) relation in the bandpasses of the SDSS system, 
which should enable the determination of more reliable distances to 
individual RR Lyrae stars for which data in the SDSS system are 
available than has been 
possible thus far. This paper presents an extension of the work by 
\citet*{mcea04}, who obtained such a PL relation for the $UBVRIJHK$ 
passbands of the Johnson-Cousins-Glass photometric system, and by 
\citet{cc08}, who studied the PL relations
in the Str\"omgren system. 
Its structure is quite similar to that in \citeauthor{mcea04} and 
\citeauthor{cc08}. 
We begin by presenting, in \S2, the theoretical framework upon which our 
study is based. In \S3, we explain the origin of the derived PL relations, 
whose calibrations are provided in \S4. Some final remarks are provided in \S6. 

\section{Models}
In order to derive the PL relations, we computed a series of horizontal 
branch (HB) simulations, following recipes similar to those presented in 
\citet{mcea04} and \citet{cc08}. We used the
evolutionary tracks given by \citet{sc98} and \citet{mcea98} 
for $Z = 0.0005,\ 0.001,\ 0.002,\ \textrm{and}\
0.006$. The main-sequence helium abundance by mass is assumed to be
$Y=0.23$, and solar-scaled compositions are adopted. 
\citeauthor{mcea04}
argue that these models are consistent with a distance modulus to the Large 
Magellanic Cloud of $(m\!-\!M)_0 = 18.47$~mag. Note, however, that this value 
is based on the empirical prescriptions for the LMC by \citet{rgea04}; using the 
independent measurements by \citet{jbea04}, a distance modulus of 
$(m\!-\!M)_0 = 18.50$~mag would derive instead. 

The mass distribution is
assumed to be a normal deviate, with a mass dispersion of $\sigma_M =
0.02$ $M_\sun$. In order to pass from the theoretical $(\log L,\,\log T_{\rm eff})$ 
plane to the empirical ones in which the magnitudes in the
SDSS photometric system ($u$, $g$, $r$, $i$, and $z$) are used, we 
have incorporated the bolometric corrections from \citet{lgea04}
to the code, over the relevant ranges of temperature and gravity. 
The blue edge of
the instability strip is computed as prescribed by \citet*{fcea87}, 
but with a shift of $-200$ K applied to the resulting
temperatures to provide better agreement with recent prescriptions
\citep{mc04}. The instability strip width is taken as $\Delta\log
T_{\rm eff} = 0.075$, which gives us the temperature of the red 
edge of the instability strip once the blue edge position has been 
found. When a star lies between the blue and red edges, 
its period is calculated based on equation~(4) of \citet{fcea98}. 
Therefore, our relations are directly applicable to fundamental-mode 
(i.e., RRab or RR0) stars, but the observed periods of first-overtone 
(RRc or RR1) stars must first be ``fundamentalized'' according to the 
relation $\log P_f = \log P_c + 0.128$ \cite[][ and references therein]{mc05}
before comparing with our results.

In this paper we are interested in finding PL relations for RR Lyrae
stars in the SDSS system. To properly take into account the impact of 
variations in HB morphology, for each of the four studied metallicities
we computed extensive series of HB simulations, including HB types that
range from very red to very blue. This leads to a total of 423,766
synthetic RR Lyrae stars that cover a wide range in metallicity and HB 
types. These stars are then used to search for possible PL relations, 
as described in the next sections. 

\begin{deluxetable}{lccclcc}
\tabletypesize{\tiny}
\tablecaption{Coefficients of the Fits}
\tablewidth{0pt}
\tablehead{
\colhead{Coefficient} & \colhead{Value} &  \colhead{Error} &&
\colhead{Coefficient} & \colhead{Value} &  \colhead{Error}}
\startdata
\multicolumn{3}{c}{$z$} && \multicolumn{3}{c}{$(u-g)_0$} \\ 
\cline{1-3}
\cline{5-7}
$a_0$ & $  1.3706$ & $0.0083$  &&  $a_0$ & $  2.1983$ & $0.0030$ \\
$a_1$ & $  0.8941$ & $0.0061$  &&  $a_1$ & $  0.6282$ & $0.0022$ \\
$a_2$ & $  0.1315$ & $0.0011$  &&  $a_2$ & $  0.0939$ & $0.0004$ \\
$b_0$ & $ -2.6907$ & $0.0372$  &&  $b_0$ & $ -1.3677$ & $0.0133$ \\
$b_1$ & $ -0.8192$ & $0.0272$  &&  $b_1$ & $ -1.0556$ & $0.0097$ \\
$b_2$ & $ -0.0664$ & $0.0049$  &&  $b_2$ & $ -0.1537$ & $0.0018$ \\
$c_0$ & $ 47.9836$ & $0.3659$  &&  $c_0$ & $ 36.8370$ & $0.1311$ \\
$c_1$ & $ 31.7879$ & $0.2676$  &&  $c_1$ & $ 24.1751$ & $0.0959$ \\
$c_2$ & $  5.2221$ & $0.0480$  &&  $c_2$ & $  3.9662$ & $0.0172$ \\
$d_0$ & $141.7704$ & $1.2837$  &&  $d_0$ & $114.0780$ & $0.4599$ \\
$d_1$ & $100.6676$ & $0.8919$  &&  $d_1$ & $ 80.9562$ & $0.3195$ \\
$d_2$ & $ 17.4277$ & $0.1541$  &&  $d_2$ & $ 13.9697$ & $0.0552$ \\
$e_0$ & $  0.3286$ & $0.0254$  &&  $e_0$ & $  1.8627$ & $0.0091$ \\
$e_1$ & $  2.0377$ & $0.0186$  &&  $e_1$ & $  1.1096$ & $0.0067$ \\
$e_2$ & $  0.3882$ & $0.0034$  &&  $e_2$ & $  0.1791$ & $0.0012$ \\
\cline{1-3}
\cline{5-7}
\multicolumn{3}{c}{$(g-r)_0$} && \multicolumn{3}{c}{$(r-i_0)$} \\ 
\cline{1-3}
\cline{5-7}
$a_0$ & $  1.1983$ & $0.0030$  &&  $a_0$ & $  0.3218$ & $0.0011$ \\
$a_1$ & $  0.6282$ & $0.0022$  &&  $a_1$ & $  0.1907$ & $0.0008$ \\
$a_2$ & $  0.0939$ & $0.0004$  &&  $a_2$ & $  0.0290$ & $0.0001$ \\
$b_0$ & $ -2.3672$ & $0.0133$  &&  $b_0$ & $ -0.8054$ & $0.0049$ \\
$b_1$ & $ -1.0552$ & $0.0097$  &&  $b_1$ & $ -0.2516$ & $0.0036$ \\
$b_2$ & $ -0.1536$ & $0.0018$  &&  $b_2$ & $ -0.0313$ & $0.0006$ \\
$c_0$ & $ 36.3361$ & $0.1311$  &&  $c_0$ & $ 15.8088$ & $0.0487$ \\
$c_1$ & $ 24.1760$ & $0.0959$  &&  $c_1$ & $ 10.8036$ & $0.0356$ \\
$c_2$ & $  3.9668$ & $0.0172$  &&  $c_2$ & $  1.8125$ & $0.0064$ \\
$d_0$ & $113.8830$ & $0.4599$  &&  $d_0$ & $ 46.0807$ & $0.1708$ \\
$d_1$ & $ 80.9365$ & $0.3195$  &&  $d_1$ & $ 33.1513$ & $0.1187$ \\
$d_2$ & $ 13.9681$ & $0.0552$  &&  $d_2$ & $  5.7868$ & $0.0205$ \\
$e_0$ & $  1.8627$ & $0.0091$  &&  $e_0$ & $  0.7327$ & $0.0034$ \\
$e_1$ & $  1.1097$ & $0.0067$  &&  $e_1$ & $  0.4298$ & $0.0025$ \\
$e_2$ & $  0.1792$ & $0.0012$  &&  $e_2$ & $  0.0700$ & $0.0004$ \\
\cline{1-3}
\cline{5-7}
\multicolumn{3}{c}{$(i-z)_0$} && &&\\
\cline{1-3}
$a_0$ & $  0.2050$ & $0.0007$  &&  &&\\ 
$a_1$ & $  0.1497$ & $0.0005$  &&  &&\\ 
$a_2$ & $  0.0242$ & $9.6205$  &&  &&\\ 
$b_0$ & $ -0.7589$ & $0.0033$  &&  &&\\ 
$b_1$ & $ -0.3685$ & $0.0024$  &&  &&\\ 
$b_2$ & $ -0.0559$ & $0.0004$  &&  &&\\ 
$c_0$ & $  8.7812$ & $0.0320$  &&  &&\\ 
$c_1$ & $  5.7009$ & $0.0234$  &&  &&\\ 
$c_2$ & $  0.9193$ & $0.0042$  &&  &&\\ 
$d_0$ & $ 32.0594$ & $0.1123$  &&  &&\\ 
$d_1$ & $ 22.1549$ & $0.0781$  &&  &&\\ 
$d_2$ & $  3.7400$ & $0.0135$  &&  &&\\ 
$e_0$ & $  0.4759$ & $0.0022$  &&  &&\\ 
$e_1$ & $  0.2728$ & $0.0016$  &&  &&\\ 
$e_2$ & $  0.0446$ & $0.0003$  &&  &&\\ 
\enddata
\label{tab:COEF}
\end{deluxetable}

\begin{deluxetable}{lccclcc}
\tabletypesize{\footnotesize}
\tablecaption{Analytical Fits: Quality Diagnostics\tablenotemark{1}}
\tablewidth{0pt}
\tablehead{
\colhead{Fit} & \colhead{$r$} &  \colhead{std. error}}
\startdata

$z$ & $0.9981$ & $0.0096$\\
$(u-g)$ & $0.9974$ & $0.0034$\\
$(g-r)$ & $0.9991$ & $0.0034$\\
$(r-i)$ & $0.9995$ & $0.0013$\\
$(i-z)$ & $0.9994$ & $0.0008$\\
\enddata
\label{tab:QUAL}
\tablenotetext{1}{For the fits given by equation~(\ref{eq:FITS}) and Table~\ref{tab:COEF}.}
\end{deluxetable}

\section{Genesis of the PL Relations in the SDSS System}

As discussed in \citet{mcea04}, the expected PL relation
must be tighter towards the redder passbands (especially in the 
near-infrared), compared to the visual bands. 
The effects of temperature and luminosity on the periods
affect strongly the shape of the resulting PL relation. In order to 
better appreciate this, recall that, from the period-mean density 
relation or Ritter's relation \citep[e.g.,][]{vab71}, periods increase
strongly with both an increase in luminosity and a {\em decrease} in
temperature. While the luminosities of RR Lyrae stars are remarkably 
uniform for a given metallicity and HB type, the introduction of 
filters, with their often strongly temperature-dependent bolometric 
corrections, may add strong slopes to the otherwise ``horizontal'' 
branch. Thus in $u$ and $g$, the cooler stars appear fainter than 
the bluer ones; conversely, in the redder passbands ($r$, $i$, and $z$), 
the cooler stars are the ones that appear brighter. Since the cooler/more 
luminous stars are the ones with longer periods, the end result is that 
the PL relation will appear increasingly tighter towards the redder 
passbands, the inverse happening towards the bluer passbands 
\citep[see][ for a detailed discussion]{mcea04}. 

This behavior is confirmed in Figure~\ref{fig:GENMAG} ({\em top row}), where
we show the changes in the absolute magnitude-log-period space, for HB
simulations computed for a rather even HB morphology, a metallicity of 
$Z=0.002$, and each of the $ugriz$ SDSS passbands. Qualitatively similar 
results are obtained for other metallicity values and HB morphologies 
as well. As can be clearly seen, it is only when the redder bandpasses 
of the SDSS sytem~-- namely, $i$ and $z$~-- are used that one begins to 
find relatively tight, simple PL relations. This behavior is in total 
agreement with the previous results by \citet{mcea04}, who had similarly
found that, in the case of the Johnson-Cousins-Glass system, such simple
PL relations are present only for $I$ and redder bandpasses.

\section{The RR Lyrae PL Relation in the SDSS System Calibrated} 

\subsection{Relations Involving a ``Pseudo-Color''} 

As shown in \citet{cc08}, the originally very poor PL relations in the 
\citet{bs63} filter system become exceedingly tight when (fairly 
reddening-independent) Str\"omgren ``pseudo-color'' 
$C_0 \equiv (u\!-\!v)_0 - (v\!-\!b)_0$ terms are 
incorporated into these relations. Can something similar be accomplished, 
in the case of the SDSS system? 

To answer this question, we have searched for a combination of blue and red 
SDSS colors that might also prove relatively reddening-free. We used the 
extinctions provided online by D. Schlegel,\footnote{{\scriptsize
{\tt http://astro.berkeley.edu/$\sim$marc/dust/data/filter.txt}}} according
to which one has $E(u)/E(\bv) = 5.16$, 
$E(g)/E(\bv) = 3.79$,
$E(r)/E(\bv) = 2.75$,
$E(i)/E(\bv) = 2.09$, and
$E(z)/E(\bv) = 1.48$. On this basis, we find that a ``pseudo-color'' defined
as 

\begin{equation}
  C_0 = (u-g)_0 - (g-r)_0 
  \label{eq:PsC}
\end{equation}

\noindent turns out to be fairly reddening-insensitive, with the unreddened 
($C_0$) and reddened ($C_1$) quantities being related by the following 
equation: 

\begin{equation}
  C_0 = C_1 - 0.32 \, E(\bv). 
  \label{eq:PsCred}
\end{equation}

That by incorporating such $C_0$-dependent terms can indeed lead to much 
tighter PL-pseudo-color (PLpsC) relations is confirmed by Figure~\ref{fig:GENMAG}
({\em bottom row}), which shows the enormous improvement over the situation 
in which no such terms are included (Fig.~\ref{fig:GENMAG}, {\em top row}). 

In the course of our research we noticed that tight 
period-{\em color}-pseudo-color (PCpsC) relations also obtain, when one includes 
such $C_0$-dependent terms. This is clearly shown in 
Figure~\ref{fig:GENCOL}.\footnote{Note that here we prefer to provide a plot 
for the $(u-z)_0$ color as opposed to $(u-g)_0$, since we have found that the 
latter presents a complex, non-linear behavior as a function of $\ln C_0$ and 
$\log P$. This is also the reason why, in equation~\ref{eq:SIMPuz} below, we 
provide a simple linar fit for $(u-z)_0$, rather than $(u-g)_0$, as a function
of $\log P$ and $\log Z$.} 
As a matter of fact, such relations are even tighter than the corresponding
PLpsC ones. Therefore, in what follows, we shall directly provide 
our theoretically calibrated PLpsC relation in a single bandpass, namely $z$
(which provides us with the highest correlation coefficient of all the 
SDSS filters), electing to provide PCpsC relations involving the remainder 
of the SDSS filters, due to their higher correlation coefficients. 
From the provided PLpsC relation in $z$ and the PCpsC 
relations, one can straightforwardly derive PLpsC relations for all other 
individual SDSS bandpasses. 

The final relations that we obtained are thus of the form: 

\begin{eqnarray}
{\rm mag \,\, or \,\,color} = \sum_{i=0}^{2} a_i (\log Z)^i & + & \sum_{i=0}^{2} b_i (\log Z)^i (\ln C_0)   \nonumber \\
															& + & \sum_{i=0}^{2} c_i (\log Z)^i (\ln C_0)^2 \nonumber \\
															& + & \sum_{i=0}^{2} d_i (\log Z)^i (\ln C_0)^3 \nonumber \\
															& + & \sum_{i=0}^{2} e_i (\log Z)^i (\log P), 
\label{eq:FITS}
\end{eqnarray}

\noindent where ``mag'' stands for the absolute magnitude in $z$, whereas
``color'' represents  
any of the colors $(u-g)_0$, $(g-r)_0$, $(r-i)_0$, and $(i-z)_0$. 
In this expression, $C_{0}$ 
is the SDSS system's pseudo-color [eq.~(\ref{eq:PsC})], and $P$ is the 
fundamentalized RR Lyrae period (in days). The corresponding coefficients,  
along with their errors, are given in Table~\ref{tab:COEF}. (Naturally, 
the $c_0$ coefficient that appears in this table is {\em not} the same 
as the pseudo-color $C_0$, which is given in capital letters 
throughout this paper to avoid confusion.)

\begin{figure*}
 \centering
  \includegraphics*[width=17cm]{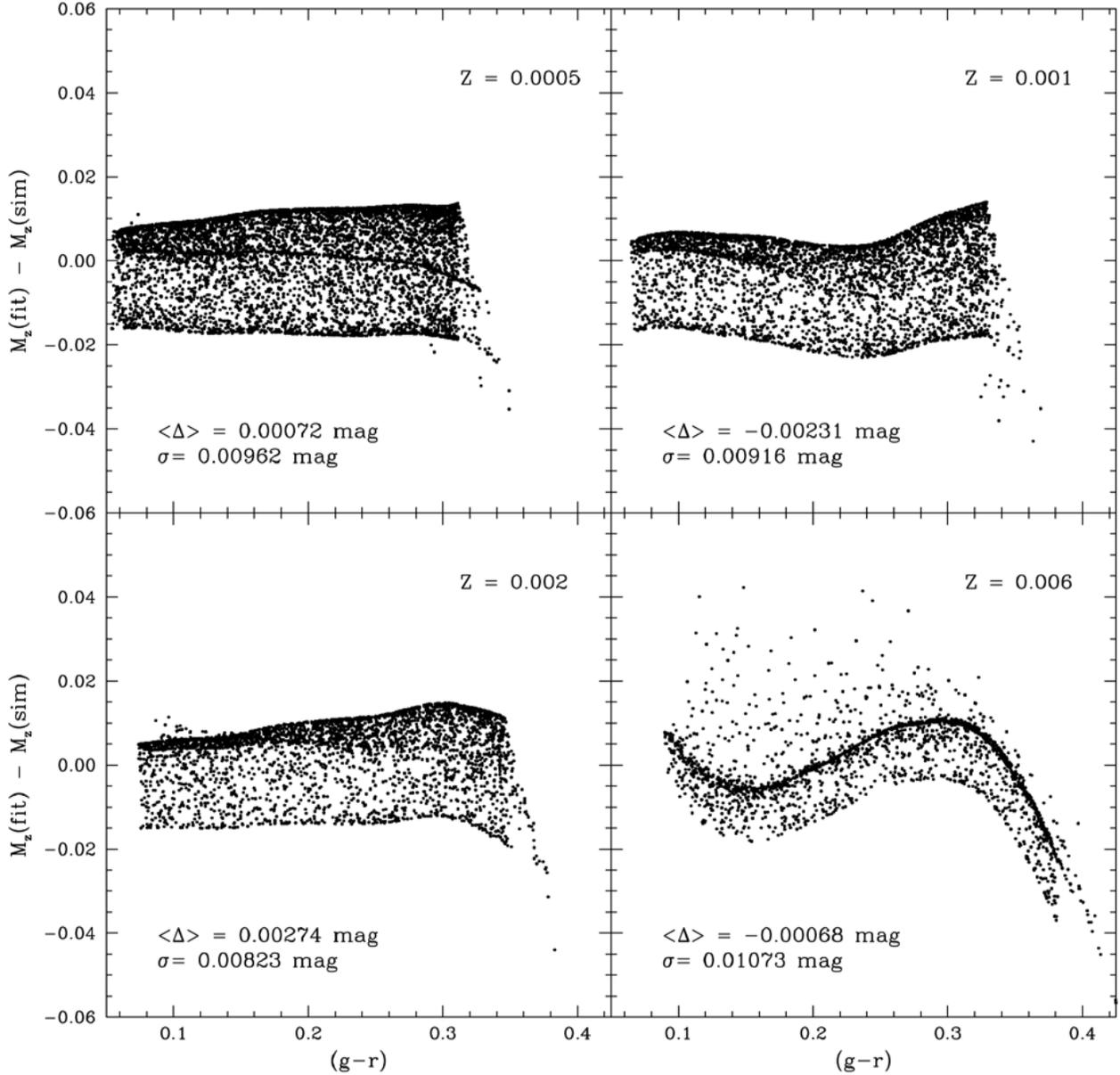}
  \caption{Difference between the absolute magnitude in $z$, as
    predicted by equation~(\ref{eq:FITS}), and its input value (from
    the HB simulations), plotted as a function of the $(g-r)_0$ color
    (from the same simulations) for four different metallicities: $Z =
    0.0005$ ({\em upper left}), 0.001 ({\em upper right}), 0.002 ({\em
      lower left}), 0.006 ({\em lower right}).  A total of 6000
    randomly selected synthetic stars is shown for all metallicities.
    The average magnitude difference is indicated in the lower left
    corner of each panel, along with the corresponding standard
    deviation. The latter is a direct indicator of the precision with
    which equation~(\ref{eq:FITS}) is able to provide the $M_z$
    values.  }
      \label{fig:RESz}
\end{figure*}

\begin{figure*}
 \centering
  \includegraphics*[width=17cm]{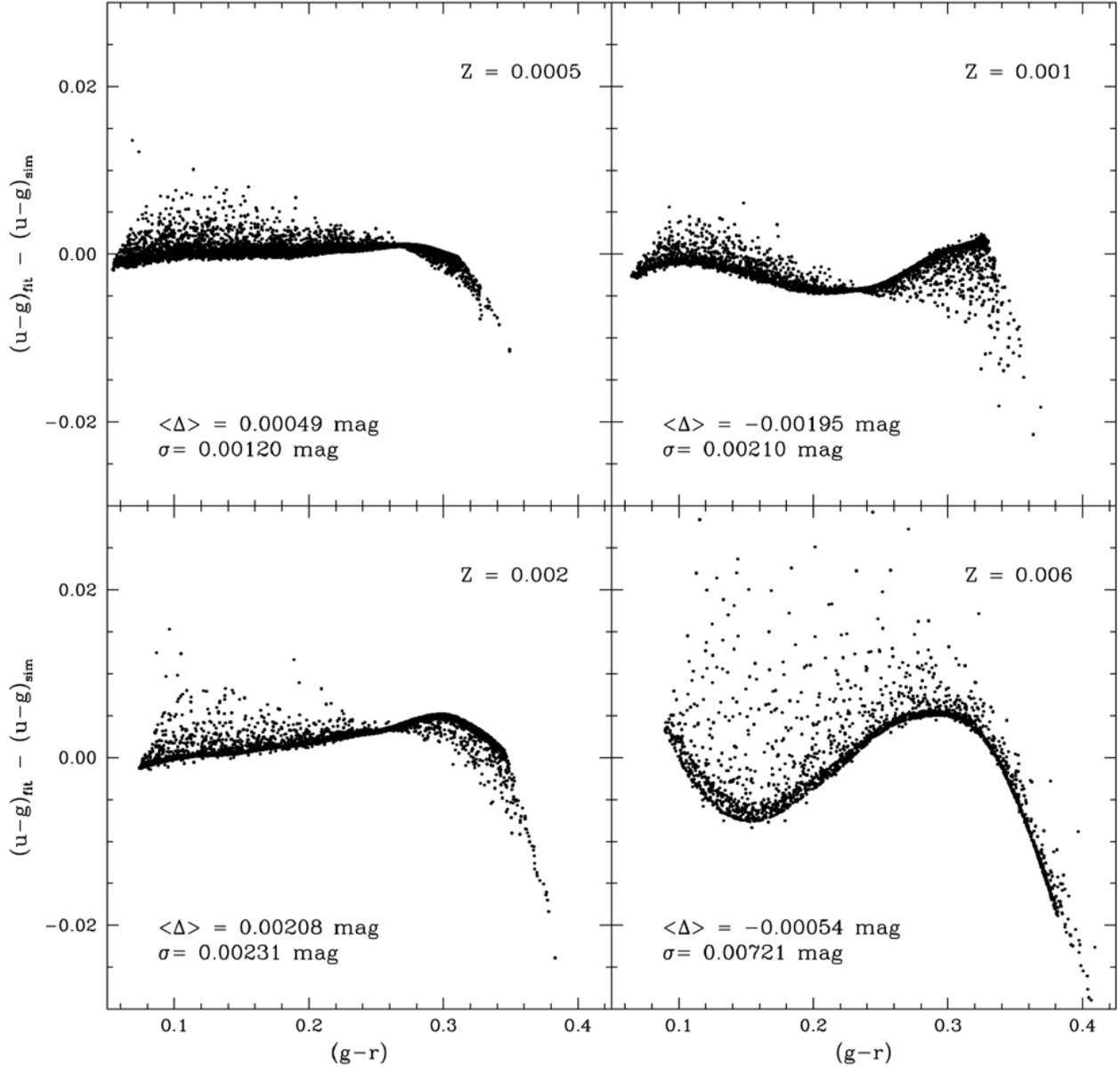}
  \caption{As in Figure~\ref{fig:RESz}, but for $(u\!-\!g)_0$.
    }
      \label{fig:RESumg}
\end{figure*}

\begin{figure*}
\centering
  \includegraphics*[width=17cm]{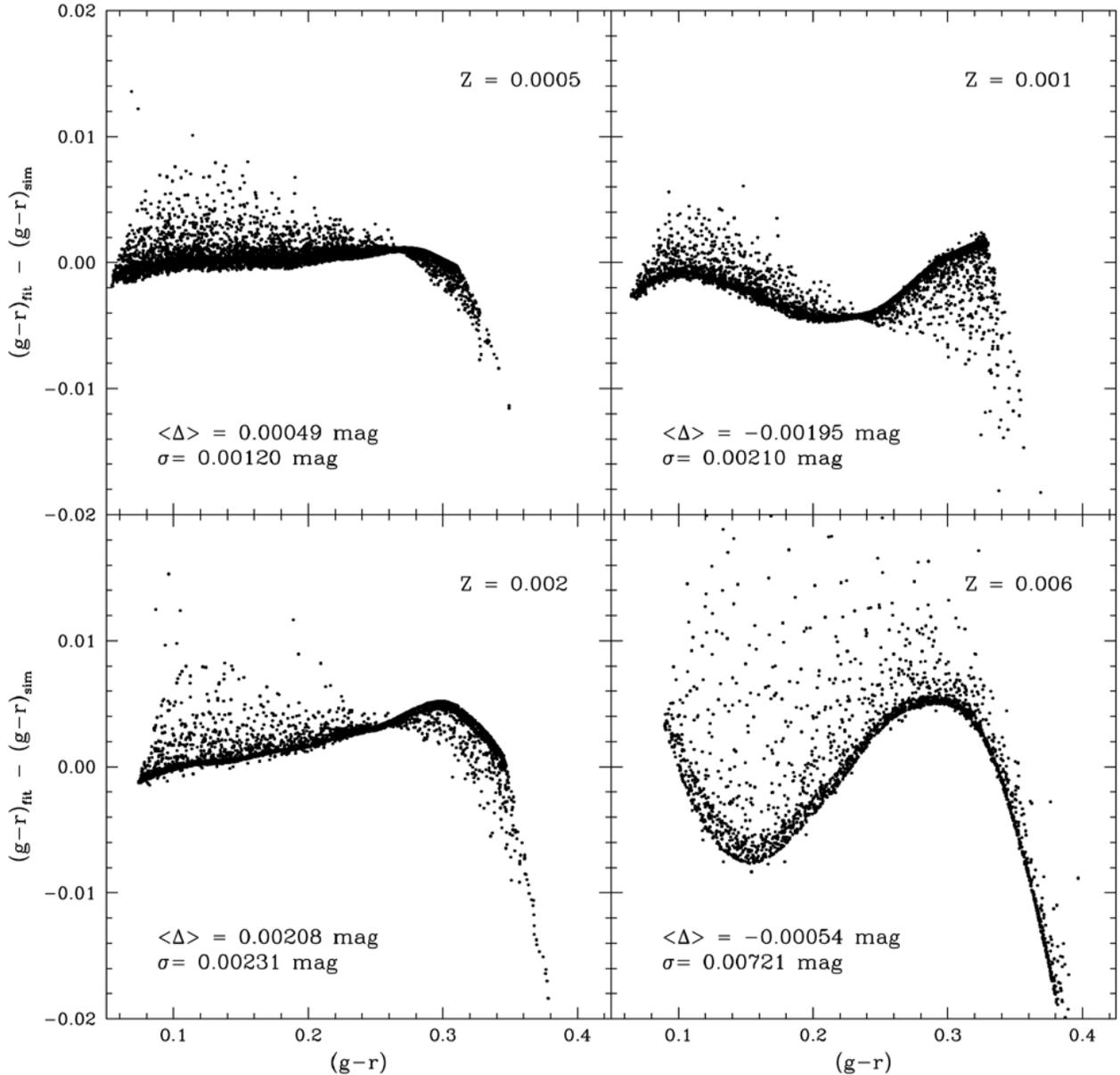}
  \caption{As in Figure~\ref{fig:RESz}, but for $(g\!-\!r)_0$.
    }
      \label{fig:RESgmr}
\end{figure*}

\begin{figure*}
\centering
  \includegraphics*[width=17cm]{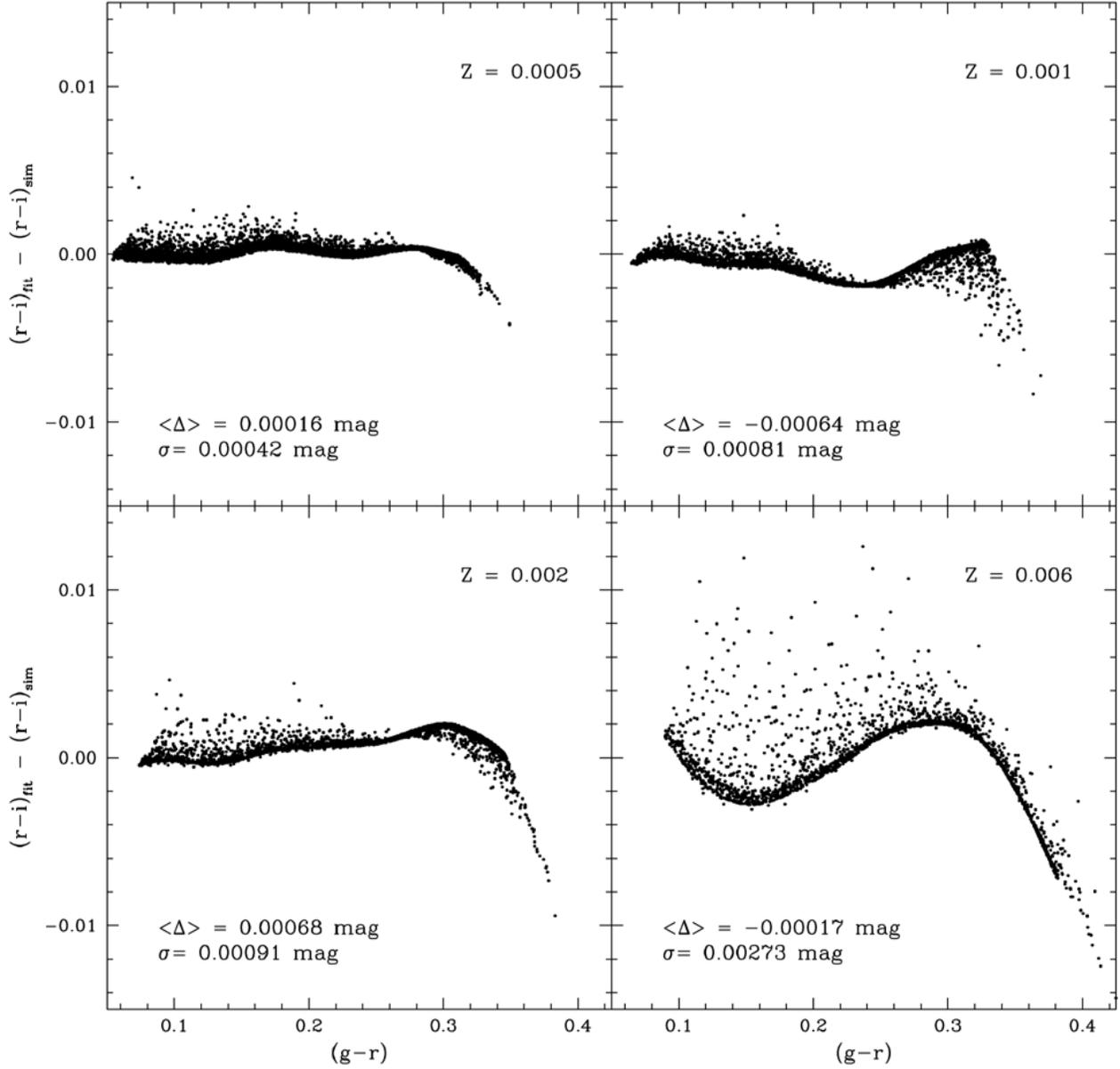}
  \caption{As in Figure~\ref{fig:RESz}, but for $(r\!-\!i)_0$.
  }
      \label{fig:RESrmi}
\end{figure*}

\begin{figure*}
\centering
  \includegraphics*[width=17cm]{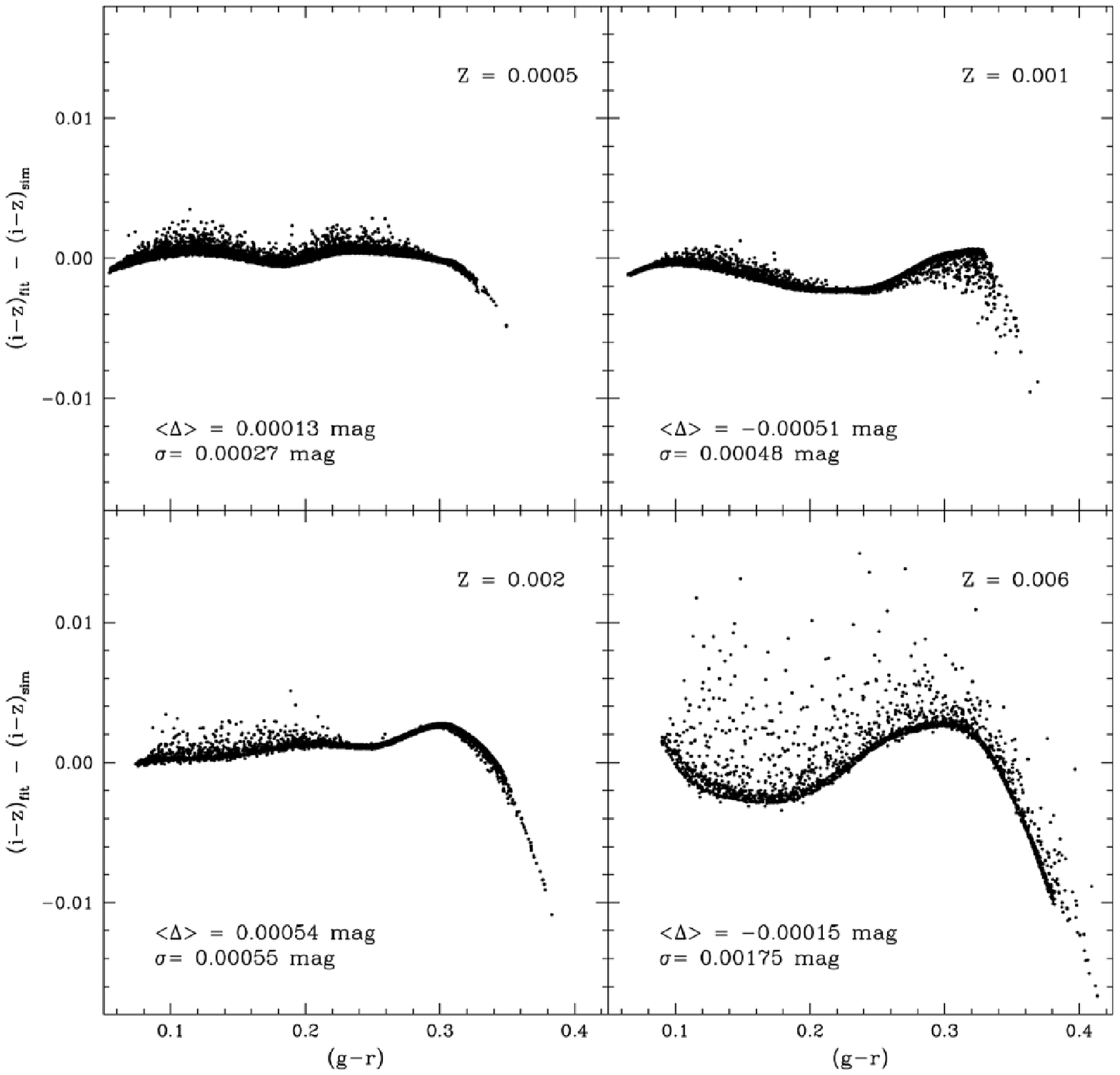}
  \caption{As in Figure~\ref{fig:RESz}, but for $(i\!-\!z)_0$.
  }
      \label{fig:RESimz}
\end{figure*}

We stress that these equations are able to reproduce the input values 
(from the HB simulations) with high precision. This is shown in 
Table~\ref{tab:QUAL}, where the correlation coefficient $r$ and the 
standard error of the estimate are given for each of the four 
equations. We also show, in Figures~\ref{fig:RESz}, \ref{fig:RESumg}, 
\ref{fig:RESgmr}, \ref{fig:RESrmi}, and \ref{fig:RESimz}, the residuals [in the sense 
eq.~(\ref{eq:FITS}) minus input values (from the simulations)] 
for a random subset of 6000 synthetic stars drawn from the original pool of 
423,766 synthetic RR Lyrae stars in the HB simulations, for the fits computed for 
$z$, $(u-g)_0$, $(g-r)_0$, $(r-i)_0$, and $(i-z)_0$, respectively. 
These plots further illustrate  
that the SDSS magnitudes and colors can be predicted from the data 
provided in equation~(\ref{eq:FITS}) and Table~\ref{tab:COEF} with a 
precision that is generally at the level of 0.01~mag (or better, 
especially at low metallicities). 

Finally, we note that equation~(\ref{eq:FITS}) can be trivially expressed
in terms of [Fe/H]; this can be accomplished using the relation 

\begin{equation}
  \log Z = {\rm [M/H]} - 1.765, 
\label{eq:LOGZ}
\end{equation} 

\noindent which is the same as equation~(9) in \citet{mcea04}. In this 
sense, the effects of an enhancement in $\alpha$-capture elements 
with respect to a solar-scaled mixture, such as observed amongst 
Galactic halo stars \citep*[e.g.,][ and references therein]{bpea05b}, 
can be taken into account by using the following scaling relation 
\citep*{msea93}: 

\begin{equation}
{\rm [M/H]} = {\rm [Fe/H]} + \log (0.638\,f + 0.362), 
\label{eq:MAUR}
\end{equation} 

\noindent where $f = 10^{\rm [\alpha/Fe]}$. However, such a relation 
should be used with due care for metallicities $Z > 0.003$ \citep{dvea00}.

\subsection{The Effect of the Helium Abundance}

\begin{figure}[t]
 \centering
  \includegraphics[width=8cm]{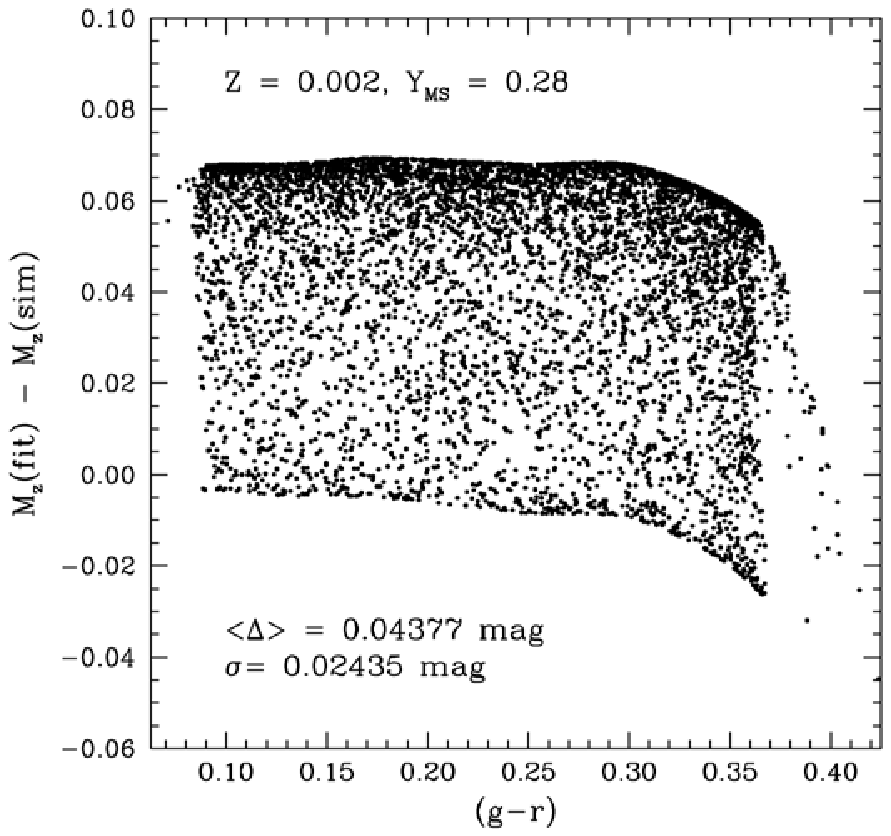}
  \caption{Effect of an increase in the helium abundance upon the
    derived PLpsC relation for $z$. Here we give the difference between the
    quantities predicted by equation~(\ref{eq:FITS}) and the input
    values from HB simulations computed for a $Y_{\rm MS} = 0.28$ and
    $Z = 0.002$, as a function of $(g\!-\!r)_0$. In this plot, 6000
    randomly selected synthetic RR Lyrae stars are shown.
	}
      \label{fig:HIYz}
\end{figure}

\begin{figure*}[t]
 \centering
  \includegraphics*[width=17cm]{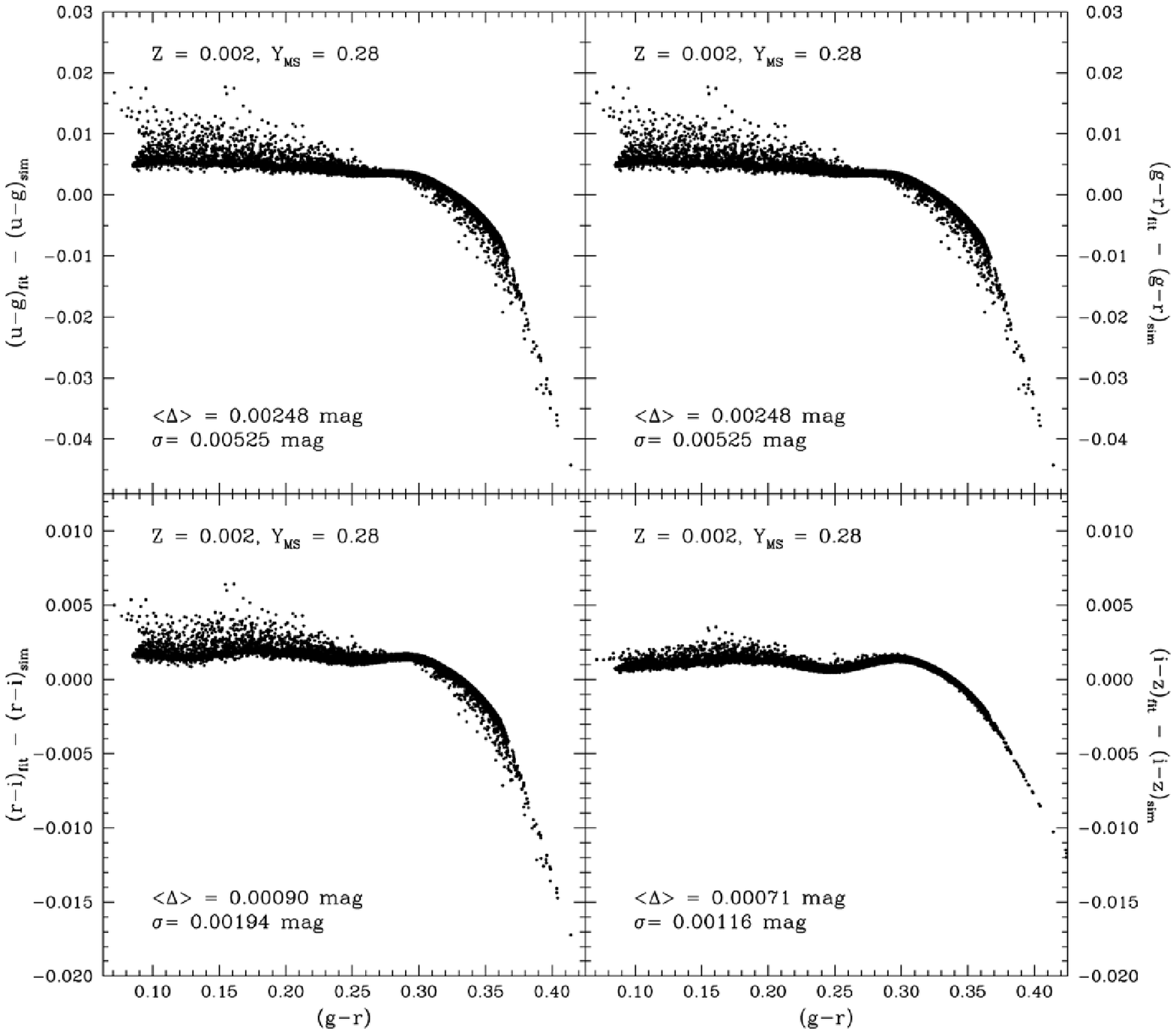}
  \caption{Effect of an increase in the helium abundance upon the
    derived relations for the $(u\!-\!z)_0$, $(g\!-\!r)_0$, $(r\!-\!i)_0$
    and $(i\!-\!z)_0$ colors. Here we give the difference between the
    quantities predicted by equation~(\ref{eq:FITS}) and the input
    value from HB simulations computed for a $Y_{\rm MS} = 0.28$ and
    $Z = 0.002$, as a function of $(g\!-\!r)_0$. In this plot, 6000
    randomly selected synthetic RR Lyrae stars are shown.
	}
      \label{fig:HIYcol}
\end{figure*}

The dependence of the RR Lyrae PLpsC relation in the SDSS system on the 
adopted width of the mass distribution, as well as on the helium 
abundance, has been analyzed by computing additional sets of synthetic 
HB's for $\sigma_M = 0.030\,M_{\odot}$ ($Z = 0.001$) and for a 
main-sequence helium abundance of 28\% ($Z = 0.002$). The effect of 
$\sigma_M$ variations was found to be negligible 
\citep[see also][ for similar results in the case of the Johnson-Cousins-Glass
and Str\"omgren systems, respectively]{mcea04,cc08}, but helium turned out
to have a more important influence (again as previously found in the other 
filter systems). The results 
are shown in Figures~\ref{fig:HIYz} (PLpsC relation in $z$) and \ref{fig:HIYcol}
(PLpsC relations). 

In the case of $M_z$, there is a clear offset in the zero point, as 
well as an increased dispersion in comparison with Figure~\ref{fig:HIYz}. 
Still, the standard deviation remains a modest $0.024$~mag~-- 
to be compared with the actual 
dispersion in $M_z$ magnitudes from the HB simulations for a fixed $Y$, 
which amounts to a full 1.25~mag (i.e., $M_z$ values for individual 
RR Lyrae stars range from 0.97~mag at their faintest to $-0.29$~mag at 
their brightest), in the $Y_{\rm MS} = 0.23$ case. 
Therefore, if a correction to the zero point $a_0$ for $M_z$ 
in Table~\ref{tab:COEF} [and also in eq.~(\ref{eq:SIMPz}) below] 
by $dM_z/dY = -0.044/(0.28-0.23) = -0.88$ 
[i.e., in the sense that eq.~(\ref{eq:FITS}) predicts too faint magnitudes] is 
duly taken into account, equation~(\ref{eq:FITS}) 
[and similarly eq.~(\ref{eq:SIMPz}) below] can also be used to provide 
useful information on the absolute magnitudes of RR Lyrae stars with enhanced  
helium abundances.

\subsection{Simple Relations} 

As in \citet{mcea04}, and for the bandpasses that show sufficiently 
tight PL relations (i.e., $i$ and $z$; see Fig.~\ref{fig:GENMAG}), 
we have computed average PL relations that
do not show an explicit dependence on $C_0$. The goal here is to 
enable an application of our derived PL relations even when 
observations in the bluer passbands of the SDSS system are not
available. We do provide, however, simple relations for {\em colors} 
involving such bluer bandpasses. The resulting relations are as follows:

\begin{equation}
M_z = 0.839 - 1.295 \, \log P + 0.211\, \log Z,
\label{eq:SIMPz}
\end{equation}

\noindent with a correlation coefficient $r = 0.97$ and a standard 
error of the estimate of 0.037~mag; 

\begin{equation}
M_i = 0.908 - 1.035 \, \log P + 0.220\, \log Z,
\label{eq:SIMPi}
\end{equation}

\noindent with a correlation coefficient $r = 0.95$ and a standard 
error of the estimate of 0.045~mag; 

\begin{equation}
(r-i)_0 = 0.184 + 0.438 \, \log P + 0.017 \, \log Z,
\label{eq:SIMPri}
\end{equation}

\noindent with a correlation coefficient $r = 0.95$ and a standard 
error of the estimate of 0.013~mag; 

\begin{equation}
(g-r)_0 = 0.640 + 0.851 \, \log P + 0.081 \, \log Z,
\label{eq:SIMPgr}
\end{equation}

\noindent with a correlation coefficient $r = 0.95$ and a standard 
error of the estimate of 0.027~mag; and 

\begin{equation}
(u-z)_0 = 2.317 + 1.472 \, \log P + 0.221 \, \log Z,
\label{eq:SIMPuz}
\end{equation}

\noindent with a correlation coefficient $r = 0.95$ and a standard 
error of the estimate of 0.045~mag. Note that, as a consequence of 
the large number of stars involved in the fits, the errors in all of 
the derived coefficient are very small (of order $10^{-4} - 10^{-3}$).
We performed tests in which quadratic terms were added to these equations, 
but in no case was the improvement of major significance, the standard
errors of the estimates generally changing only in the third decimal
place.

\subsection{On Applying Our Relations to RR Lyrae Stars}\label{sec:CAVEATS} 

When applying our equation~(\ref{eq:FITS}) in globular cluster work, 
the metallicity of the cluster will often be known a priori. However, 
metallicity estimates may also be unavailable, especially when dealing 
with field RR Lyrae stars. Yet, for a reliable application of our relations 
to field stars, an estimate of their metallicities must be provided. 

The SDSS system itself may itself come to our rescue in such a case. 
We recall that estimates of the RR Lyrae 
metallicities can be obtained on the basis of their $V$-band light curves 
using Fourier decomposition \citep*[e.g.,][]{jk96,jj98,kk07,smea07}. 
Transformation equations between the SDSS system and the Johnson-Cousins
system have been provided in the literature \citep*[e.g.,][]{skea05}, and an 
updated list of such transformation equations is maintained at the 
SDSS web site.\footnote{
{\scriptsize{\tt http://www.sdss.org/dr4/algorithms/sdssUBVRITransform.html}}}
For instance, from the current ``Lupton set'' one finds the following: 

\begin{equation}
  V = g - 0.2906 \, (u-g) + 0.0885,
  \label{eq:LUP1}
\end{equation}

\noindent with a $\sigma = 0.013$~mag, and 

\begin{equation}
  V = g - 0.5784 \, (g-r) - 0.0038,
  \label{eq:LUP2}
\end{equation}

\noindent with a $\sigma = 0.005$~mag. Thus, on the basis of $V$-band light 
curves computed from SDSS $u$- and $g$- (or, alternatively, $g$- and $r$-) 
band magnitudes, one should be able to estimate metallicities for individual 
RR Lyrae stars, through Fourier decomposition. We note, in addition, that in 
a forthcoming paper (Catelan \& C\'aceres 2008, in preparation) we will be 
providing analytical fits that should allow one to estimate metallicity 
values directly from the SDSS photometry.

The reader should be warned that our relations should be compared 
against empirical quantities obtained for the so-called {\em equivalent static 
star}. Several procedures have been advanced in the literature for the 
determination of the latter on the basis of empirically derived magnitudes 
and colors 
\citep*[e.g.,][ and references therein]{gbea95}. In particular, one should note 
that, according to the hydrodynamical models provided by \citeauthor{gbea95}, 
one should expect differences between temperatures derived from  
intensity- or magnitude-averaged colors, on the one hand, and those based 
on the actual color of the equivalent static star, on the other. As a workaround, 
these authors set forth very useful amplitude-dependent corrections, which become 
more important the bluer the bandpass. More recently, \citet{mmea06} analyzed
the problem in the specific case of the SDSS system, concluding that  
intensity averages, though not perfect, are to be preferred over averages
carried out in magnitude units. 
Unfortunately, tables with amplitude-dependent corrections that 
would allow one to properly compute magnitudes and colors in the SDSS system 
for the equivalent static star 
\citep[i.e., similar to those provided by][ in the Johnson-Cousins sytem]{gbea95} 
have not yet been provided in the literature.  
One way or another, the reader should note that $C_0$, 
being a difference between two colors, 
is presumably affected to a lesser degree than are the colors themselves
\citep[see also][]{cc08}. Needless to say, 
observers are also strongly warned against using {\em single-epoch}  
photometry to derive $C_1$ values to be used along with our relations.

\section{Summary}
We have provided the first extensive calibration of the RR Lyrae PL 
(and PC) relations in the SDSS $ugriz$ filter system. As in \citet{mcea04}, 
we find that these PL relations become progressively tighter for the 
redder passbands, those in $i$ and $z$ appearing particularly 
promising. We provide very precise relations involving a newly defined, 
fairly reddening-insensitive pseudo-color $C_0 \equiv (u-g)_0 - (g-r)_0$. 
$C_0$-independent, though less precise, average relations are also provided 
for those cases in which observations in all five SDSS filters may not be 
available. Our relations should be especially useful for the calculation 
of distances and reddenings to even individual field RR Lyrae stars.

\acknowledgments We thank H. A. Smith and N. De Lee for interesting 
discussions. The authors acknowledge financial support by Proyecto 
FONDECYT Regular No.~1071002.

{}


\begin{thebibliography}{}

\bibitem[Abadi et al.(2003)]{maea03}
  Abadi, M. G., Navarro, J. F., Steinmetz, M., \& Eke, V. R. 2003, \apj, 
    591, 499

\bibitem[Belokurov et al.(2006)]{vbea06} 
  Belokurov, V., et al. 2006, \apj, 654, 897

\bibitem[Bono et al.(1994)Bono, Caputo, \& Stellingwerf]{gbea94} 
  Bono, G., Caputo, F., \& Stellingwerf, R.~F.\ 1994, \apjl, 432, L51 

\bibitem[Bono et al.(1995)Bono, Caputo, \& Stellingwerf]{gbea95}
  Bono, G., Caputo, F., \& Stellingwerf, R. F. 1995, \apjs, 99, 263

\bibitem[Borissova et al.(2004)]{jbea04}
  Borissova, J., Minniti, D., Rejkuba, M., Alves, D., Cook, K. H., \& Freeman, 
    K. C. 2004, \aap, 423, 97

\bibitem[Caputo et al.(1987)]{fcea87} 
  Caputo, F., De Stefanis, P., Paez, E., \& Quarta, M. L. 1987, \aaps, 
    68, 119 

\bibitem[Caputo et al.(1998)Caputo, Marconi, \& Santolamazza]{fcea98} 
  Caputo, F., Marconi, M., \& Santolamazza, P. 1998, \mnras, 293, 364

\bibitem[Catelan(2004)]{mc04} 
  Catelan, M.\ 2004, \apj, 600, 409 

\bibitem[Catelan(2005)]{mc05} 
  Catelan, M.\ 2005, preprint (astro-ph/0507464)

\bibitem[Catelan et al.(1998)]{mcea98} 
  Catelan, M., Borissova, J., Sweigart, A.~V., \& Spassova, N.\ 1998, \apj, 
    494, 265 

\bibitem[Catelan et al.(2004)Catelan, Pritzl, \& Smith]{mcea04} 
  Catelan, M., Pritzl, B.~J., \& Smith, H.~A.\ 2004, \apjs, 154, 633 

\bibitem[Cort\'es \& Catelan(2008)]{cc08} 
  Cort\'es, C., \& Catelan, M.\ 2008, \apjs, in press (astro-ph/0802.2309)

\bibitem[De Lee et al.(2007)De Lee, Smith, \& Beers]{ndlea07} 
  De Lee, N. M., Smith, H. A., \& Beers, T. C. 2007, BAAS, 39, 211.2502

\bibitem[Girardi et al.(2004)]{lgea04} 
  Girardi, L., Grebel, E. K., Odenkirchen, M., \& Chiosi, C. 2004, \aap,
    422, 205

\bibitem[Gratton et al.(2004)]{rgea04}
  Gratton, R. G., Bragaglia, A., Clementini, G., Carretta, E., Di Fabrizio, L., 
    Maio, M., \& Taribello, E. 2004, \aap, 421, 937

\bibitem[Greco et al.(2008)]{cgea08}
  Greco, C., et al. 2008, \apjl, 675, L73	
	
\bibitem[Jurcsik(1998)]{jj98}
  Jurcsik, J. 1998, \aap, 333, 571

\bibitem[Jurcsik \& Kov\'acs(1996)]{jk96}
  Jurcsik, J., \& Kov\'acs, G. 1996, \aap, 312, 111

\bibitem[Karaali et al.(2005)Karaali, Bilir, \& Tun\c{c}el]{skea05}  
  Karaali, S., Bilir, S., \& Tun\c{c}el, S. 2005, PASA, 22, 24
  
\bibitem[Kov\'acs \& Kupi(2007)]{kk07}
  Kov\'acs, G., \& Kupi, G. 2007, \aap, 462, 1007

\bibitem[Kuehn et al.(2008)]{ckea08}
  Kuehn, C., et al. 2008, \apjl, 674, L81  

\bibitem[Marconi et al.(2006)]{mmea06}
  Marconi, M., Cignoni, M., Di Criscienzo, M., Ripepi, V., Castelli, F., 
    Musella, I., \& Ruoppo, A. 2006, \mnras, 371, 1503
  
\bibitem[Morgan et al.(2007)Morgan, Wahl, \& Wieckhorst]{smea07}
  Morgan, S. M., Wahl, J. N., \& Wieckhorst, R. M. 2007, \mnras, 374, 1421

\bibitem[Pritzl et al.(2005)Pritzl, Venn, \& Irwin]{bpea05b}
  Pritzl, B. J., Venn, K. A., \& Irwin, M. 2005, \aj, 130, 2140
  
\bibitem[Salaris et al.(1993)Salaris, Chieffi, \& Straniero]{msea93} 
  Salaris, M., Chieffi, A., \& Straniero, O.\ 1993, \apj, 414, 580 

\bibitem[Sesar et al.(2007)]{bsea07}
  Sesar, B., et al. 2007, \aj, 134, 2236  
  
\bibitem[Str{\"o}mgren(1963)]{bs63} 
  Str{\"o}mgren, B.\ 1963, \qjras, 4, 8 

\bibitem[Sweigart \& Catelan(1998)]{sc98}
  Sweigart, A. V., \& Catelan, M. 1998, \apjl, 501, L63

\bibitem[van Albada \& Baker(1971)]{vab71} 
  van Albada, T. S., \& Baker, N. 1971, \apj, 169, 311

\bibitem[VandenBerg et al.(2000)]{dvea00} 
  VandenBerg, D.~A., Swenson, F.~J., Rogers, F.~J., Iglesias, C.~A., \& Alexander, 
    D.~R.\ 2000, \apj, 532, 430 

\bibitem[Wilhelm et al.(2008)]{rwea08}
  Wilhelm, R., et al. 2008, preprint (astro-ph/0712.0776)	
	
\end{thebibliography}
\end{document}